\documentclass[12pt,a4paper]{article}
\usepackage{lscape}
\usepackage{graphicx}

\begin{document}
\title{ Disordered systems on various time scales: a-Si$_3$B$_3$N$_7$ and homogeneous sintering
}
\author{ J. C. Sch{\"o}n$^a$, A. Hannemann$^a$, G. Sethi$^a$, M. Jansen$^a$, \\
P. Salamon$^b$, R. Frost$^c$, L. Kjeldgaard$^d$ \\ $^a$Max-Planck-Institut 
f{\"u}r Festk{\"o}rperforschung\\
 Heisenbergstr. 1, D-70569 Stuttgart, Germany \\
$^b$Dept.\ of Mathematics, San Diego State University,
San Diego, USA \\
$^c$San Diego Supercomputer Center,
San Diego, USA  \\
$^d$MAX-Laboratory,
Lund, Sweden
} 
\date{\today}
\maketitle
\renewcommand{\baselinestretch}{1.25}
\small
\normalsize
\section{Introduction} \label{introduction}
Classically, the properties and dynamics of materials systems can, in principle, be described with the help of their energy landscape. At each given moment in time, we associate with the current configuration of N particles a point in a 3N-dimensional space and an energy. The dynamics of the system is reflected in the movement of this point on the energy landscape. However, this energy landscape is highly complicated, and trying to analytically compute the dynamics is not possible. Thus we are thrown back to the use of numerical simulations, which gravely limit the time window over which we are able to perform accurate calculations. Furthermore, it can be rather difficult to extract physically relevant results from large multi-particle simulations.

However, the systems often possess features that can aid our quest. Many properties of a given system do not depend on time or change according to simple (phenomenological) laws, at least within the accuracy of our measurements. This allows us to understand the behavior of the system quite well by using simple (phenomenological) models. Frequently, the underlying reason for this success is the applicability of separation of time scales arguments that one can use to distill the important degrees of freedom relevant on short or long time scales. 

Some of the typical processes, each possessing characteristic time scales, relevant in materials systems are e.g.\
\begin{itemize}
\item Relaxation of electronic degrees of freedom
\item Lattice vibrations (phonons)
\item Breaking/formation of chemical bonds
\item Changes in the structure of sub-systems, switching between modifications, growth of nuclei
\item Grain boundary movement / annealing
\item Diffusional transport, and relaxation to equilibrium due to diffusion
\item Decay of non-equilibrium phases
\item Appearance of fluctuations of certain size (formation of nuclei)
\item Transitions among double (multi) well systems
\item Equilibration of various physical properties
\item Changes in the shapes of sub-systems
\item Interactions among sub-systems
\item $\ldots$
\end{itemize}

After a short general discussion concerning simplifications of dynamics from the point of view of time scales, we will present examples of relevant time scales and their application to modeling for two systems, amorphous silicon boron nitride, a-Si$_3$B$_3$N$_7$, and homogeneous sinters.

\section{Time scales and modeling}
Materials systems consisting of many atoms, are characterized by a very large number of degrees of freedom $(x_k) = (x_1,\ldots,x_N)$ ($N \approx 10^{23}$). As a consequence, a complete description of their dynamical properties by solving the appropriate quantum mechanical equations over the many time scales of interest constitutes an impossible task. 

However, once one is willing to restrict oneself to particular questions and properties concerning such a system, it is in many cases feasible to construct a simplified model of the system whose dynamics is amenable to numerical or (ideally) analytical solution. Technically, such simplifications can be achieved by a number of approaches:
\begin{itemize}
\item Direct reduction in the number of degrees of freedom: 
\begin{equation}
(x_1,\ldots,x_N) \rightarrow (x_1,x_2).
\end{equation} 
Such an approach is particularly useful, if we are interested in a limited time range, over which we want to compute the dynamics, which only depends on the variables $(x_1,x_2)$ in that time-range. If $(x_1,x_2)$ and $(x_3,\ldots,x_N)$ are fast and slowly varying variables, respectively, we can replace $(x_3,\ldots,x_N)$ by some constant initial value $(x_3,\ldots,x_N) \approx (x_3(0),\ldots,x_N(0))$. On the other hand, if $(x_1,x_2)$ vary slowly, while $(x_3,\ldots,x_N)$ fast fluctuate around some average value, it is appropriate to set the latter equal to this average $(x_3,\ldots,x_N) \approx (x_3^{(av)},\ldots,x_N^{(av)})$.

Mathematically, this kind of analysis can be performed in a formal way, and we refer to the literature on differential equations for more detail\cite{Bender78a}. 

\item Simplification of the dynamics, up to the point where analytically solvable equations for the observables of interest, $[A_s]$, result (on a restricted time scale), 
\begin{equation}
[A_s(x_1,\ldots,x_N)] \rightarrow [\hat{A}_s(x_1,\ldots,x_N)].
\end{equation} 
This is a very common approach, which consists of applying physically reasonable approximations not to the set of variables but to the dynamical equations instead.

Often, one can combine this approach with the one mentioned above, and, in addition, reduce the number of variables one considers, $[\hat{A}_s(x_1,\ldots,x_N)] \approx [\hat{A}_s(x_1,x_2)]$

\item A related approach consists of focussing the analysis on a small number of quantities of interest $[A_s(x_1,\ldots,x_N)]$, which might exhibit a simpler dynamics, possibly depending on fewer variables, than the full system with all observables $[A_s(x_1,\ldots,x_N)]$, $s = 1,\ldots,m$: 
\begin{equation}
[A_s(x_1,\ldots,x_N)] \rightarrow  [A_1(x_1,\ldots,x_N),A_2(x_1,\ldots,x_N)] \rightarrow [A_1(x_1,x_2),A_2(x_1,x_2)].
\end{equation}

\item Finally, and this constitutes perhaps the crowning achievement in a separation of time scales approach, one can set up a hierarchy of dynamical systems in a stepping-stone procedure. Starting with the full system with all degrees of freedom on the fastest time scale $(1)$, $[A_s^{(1)}(x_k^{(1)})]$, one successively reduces the number of degrees of freedom until one derives a (phenomenological) set of equations for those observables that are relevant for the dynamical description on the longest times, up to infinity.

One should note that during this procedure, we might employ different sets of observables $[A_s^{(i)}]$ and variables $(x_k^{(i)})$ for the description of the dynamics on different time scales $(i)$. Furthermore, in many cases, the dynamics on the longer time scale will depend on parameters $\alpha$ that incorporate the outcome of the fast dynamics, such as a diffusion constant computed from an atomic simulation. In general, these parameters $\alpha$ will parametrically depend on the values of the slow variables as prescribed for the fast-scale calculation: 
\begin{equation}
[A_s^{(i)}(x_k^{(i)})] \rightarrow [A_s^{(i+1)}(x_k^{(i+1)};\alpha(x_k^{(i+1)}))].
\end{equation} 
\end{itemize}

In many cases, a rigorous mathematical analysis of the approximations involved is not possible or insufficient, and needs to be supplemented by physical reasoning born from experience. For materials systems, three general concepts have turned out to be particularly helpful, partial (themodynamical) equilibrium, local (thermodynamical) equilibrium, and stationary state approximations.

\begin{itemize}
\item Partial equilibrium describes a situation, where the time scale of some degrees of freedom is so short that they de-couple from the rest of the system in the following sense: Before the slow degrees of freedom change appreciably, the fast ones reach thermodynamical (statistical) equilibrium among themselves. This allows two kinds of simplifications: 

On very short time scales, we can study the relaxation behavior of the fast degrees of freedom while keeping the slow degrees of freedom fixed. 

Conversely, one can study the long-term dynamics of the slow degrees of freedom by replacing the fast degrees of freedom by their statistical averages. Such averages might depend on the current values of the slow variables, and thus require the introduction of additional (phenomenological) terms in the dynamical equations for the slow variables on these variables via 
\begin{eqnarray}
\frac{dx_i}{dt} =F(x_1,x_2,x_3,\ldots,x_N), i=1,\ldots,N \rightarrow \nonumber \\
\frac{dx_i}{dt} \approx F(x_1,x_2,x_3^{av}(x_1,x_2),\ldots,x_N^{av}(x_1,x_2)), i = 1,2.
\end{eqnarray} 
Furthermore, in many situations, one needs to include the fluctuations in the fast variables (around the respective averages) in the dynamics for the slow variables. One common way to achieve this is by adding the fluctuations as a stochastic force acting on the slow degrees of freedom.\cite{Lifshitz81a} 

\item A second, related type of simplifications derives from the concept of local (thermodynamic) equilibrium. This refers to the equilibration of all (!) degrees of freedom on the (3N-dimensional) energy landscape of the system. As a consequence, the energy landscape can be divided into a set of locally ergodic regions $\mathcal{R}$, where the system can be considered being in (metastable) thermodynamic equilibrium\cite{Schoen98e,Schoen01d}. However, in contrast to the partial equilibrium, this state of affairs only persists for a limited time, until the escape time $\tau_{esc}$ from the locally equilibrated regions $\mathcal{R}$ has been reached. Thus, the local ergodicity of $\mathcal{R}$ is only present on a certain time scale, 
\begin{equation}
\tau_{eq} < t_{obs} \ll \tau_{esc},
\end{equation} 
where $\tau_{eq}$ indicates the time required for the regions $\mathcal{R}$ to equilibrate. Mathematically, $\tau_{eq}$ is defined as the smallest time $t'$ for which the difference between time- and ensemble average of observables in $\mathcal{R}$ falls below a certain prescribed accuracy $a$ for all $t > t'$,
\begin{equation}
|\langle \mathcal{O}\rangle_{ens} - \langle \mathcal{O}\rangle_{t}| < a \quad \forall t > t' \ge \tau_{eq}.
\end{equation} 
For $t_{obs} > \tau_{esc}$, the regions $\mathcal{R}$ lose their special relevance. Usually, some larger regions $\mathcal{R'}$ take their place, which are locally ergodic for times between $\tau_{eq}(\mathcal{R'}) \ge \tau_{esc}(\mathcal{R})$ and $\tau_{esc}(\mathcal{R'})$.

From the point of view of modeling the dynamics, the relevant configuration space can be viewed as being essentially discretized for time scales $t_{obs} > \tau_{eq}(\mathcal{R})$. Here, the regions $\mathcal{R}$ correspond to points on a metagraph (decribed via variables $(y_j)$), on which the dynamics of the (possibly new) observables $[A_s^{(i+1)}]$ proceeds, e.g.:
\begin{equation}
[A_s^{(i)}(x_1,\ldots,x_N)] \rightarrow [A_s^{(i+1)}(y_1,y_2)].
\end{equation}
In certain situations, one can treat $(y_j)$ as if it were a set of continuous variables. Describing the dynamics commonly requires 
developing phenomenological deterministic or stochastic models for $(y_j)$ and/or $[A_s^{(i+1)}]$.

In many situations, we encounter a combination of partial and local equilibrium. This commonly occurs, if we are dealing with local spatial equilibrium, where the system can be modeled as a union of spatially separate pieces, with each being in local equilibrium. 

\item A special instance of local/partial equilibration that allows the use of separation of time scales arguments is the so-called steady-state-situation. Here, external forces are applied to the system, such that essentially constant matter/energy flows are created in the system. Thus, the system never reaches a "traditional" thermodynamic equilibrium, since e.g.\ the temperature is no longer constant throughout the system. However, we might be able to establish a time-independent temperature distribution throughout the material, which would correspond to a locally/partially equilibrated state under certain boundary conditions. This allows us to separate slow and fast degrees of freedom, where the fast ones equilibrate within a short time, and the relevant dynamics of the system on long time scales can be modeled using only a small number of variables.

\end{itemize}

\section{Amorphous Silicon Boron Nitride a-Si$_3$B$_3$N$_7$}
In an amorphous system such as a-Si$_3$B$_3$N$_7$, processes take place over a range of very different time scales. Many of these are also present in typical crystalline solids, but the fact that amorphous systems do not reside in the region around the global (or 'for all practical purposes' global) minimum of the energy landscape leads to a steady drift in configuration space on all time scales. Although local ergodicity is usually present on certain time scales, we always find an escape time beyond which the system must be treated as unstable, a process which is commonly denoted as aging. In contrast, systems that explore the neighborhood of the global minimum (or any minimum with a similar barrier structure representing a stable crystalline modification) of the landscape usually possess a limiting time scale, beyond which the system can be treated as globally ergodic, i.e. as being in global equilibrium. 

Nevertheless, simplified modeling of amorphous systems using separation of time scales is clearly possible. In the following, we are going to list some of the most important time scales (with estimates of their value) and discuss their use in modeling the dynamics of a-Si$_3$B$_3$N$_7$.
\begin{itemize}
\item Movement and relaxation of electronic degrees of freedom ($\tau \approx 10^{-15}$ sec). \\
This is the time scale on which electronic, magnetic or optical properties, need to be addressed. From the point of view of a stepping stone approach for the longer time dynamics, one would attempt to employ ab initio calculations to derive empirical potentials that can be used in subsequent steps when performing Monte Carlo or Molecular Dynamics calculations. For a-Si$_3$B$_3$N$_7$, several such potentials have been constructed\cite{Marian00a}, which have been employed in the calculations discussed below.
\item Lattice vibrations ($10^{-9}$ sec $ > \tau^{vib} > 10^{-14}$ sec) \\
We note that if we perform simulations of the system in the picosecond region, we are able to study the vibrations of the network, because the time scales on which the network itself changes ($\tau_{BSP}$, see below) are longer by far, at least at low temperatures. This can be seen in the vibrational spectra along the trajectories (c.f.\ fig.\ \ref{phonon1}). 
\begin{figure}
\centering
\includegraphics[width=0.6\textwidth,angle=0]{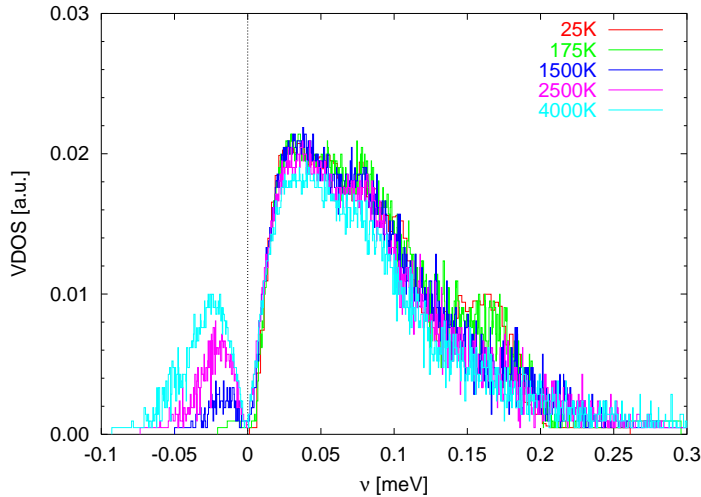}
\caption{Typical vibrational spectrum along trajectories for five different temperatures between $ 25$ K and $ 4000$ K. 'Negative' frequencies $\omega$ correspond to imaginary modes.
\label{phonon1}} 
\end{figure}
There, we observe that for low temperatures, only a very small percentage indicate imaginary modes, i.e., that the trajectory is in a neighborhood of a saddle point in the corresponding eigendirection. Clearly, most of the time we are in the process of vibrating around a local minimum. Only at high temperatures close to or above the glass transition point ($T_c$, see below) do we encounter a substantial fraction of imaginary modes.\cite{Hannemann02b}

Encouraged by the separation of time scales argument, one can study the spectrum of real eigenvalues of the minima located along the actual trajectories (c.f.\ fig.\ \ref{phonon2}). 
\begin{figure}
\centering
\includegraphics[width=0.6\textwidth,angle=0]{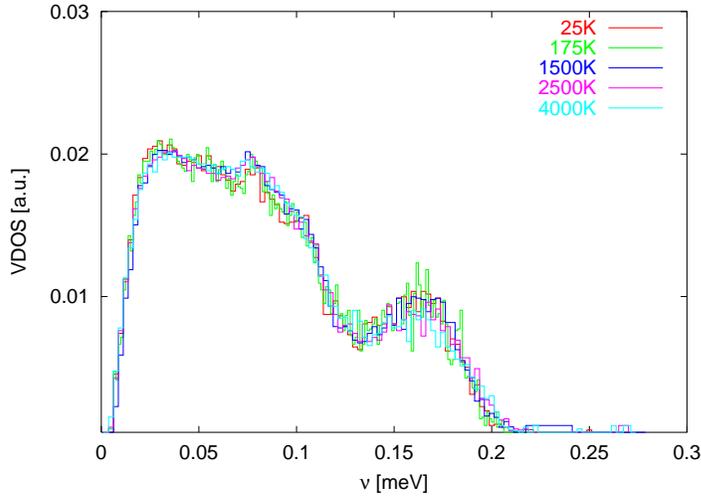}
\caption{Typical vibrational spectrum for local minima of the energy landscape observed along trajectories for five different temperatures between $ 25$ K and $ 4000$ K.
\label{phonon2}} 
\end{figure}
Here, we find that the energy landscape of the liquid ($T > T_c$) and the glassy  ($T < T_c$) state are very similar with regard to the local minima lying below. The general shape of the frequency spectrum of all minima is nearly identical, and one only observes a slight change towards a higher average frequency at minima with lower potential energy. Since the deeper-lying local minima are encountered more frequently in the glassy state, there should therefore occur a slight softening of the vibrations when passing through the glass transition region upon increasing the temperature. But we note that this effect is not large enough to account for the peak in the specific heat (c.f.\ fig.\ \ref{specheat}) observed in this temperature range\cite{Hannemann02a}. Regarding this figure, one should note that the contribution of the kinetic energy ($3/2k_B$) has already been subtracted. For $T \rightarrow \infty$, $C_V$ drops below $3/2k_B$ because the anharmonic part of the potential becomes dominant and the potential goes to zero for long distances. For $T < T_c$, $C_V$ falls below $3/2k_B$ because of aging effects involved in the numerical computation of $\langle E \rangle (T)$ for neighboring temperatures $T_{1,2}$ ($\Delta T = T_2 - T_1$). This effect is a consequence of the different rates of decrease $m$ in minimum energies with temperature shown in fig.\ \ref{epot}, which have a maximum around $1750$ K.
\begin{figure}
\centering
\includegraphics[width=0.5\textwidth,angle=0]{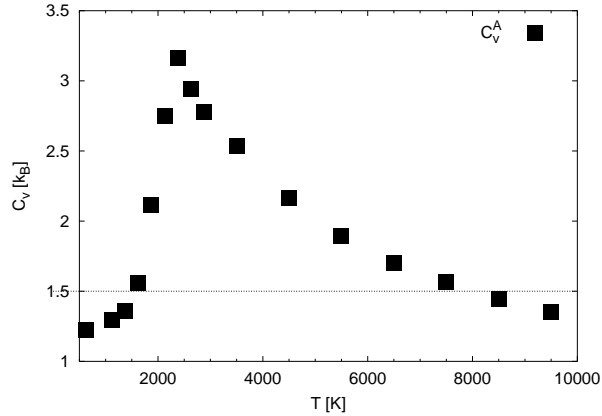}
\caption{Specific heat per atom calculated via $C_V = \Delta \langle E \rangle / \Delta T$. 
\label{specheat}} 
\end{figure}
\item Time scales associated with the transport of energy \\
A typical time scale of energy transport that is relevant during molecular dynamics simulations is given by: $\tau_S \propto L/v_S \approx 10^{-11} - 10^{-13}$ sec, where $L$ is the length of the simulation cell (typically, $L \approx 10 - 100$ {\AA}), and $v_S$ is the speed of sound (typically, $v_S \approx 10^3 - 10^4$ m/sec). Turning this relation around, we can estimate the speed of sound in our system by studying the time evolution of a heat pulse that is excited along one layer of the system at a given time. From the speed with which the energy spike travels through the system, we can deduce a value for $v_S \approx 3-4 \times 10^{3}$ m/sec for the energy flow\cite{Sethi02a}, a value which lies in the range of the speeds of sound found experimentally in crystalline Si$_3$N$_4$ ($v_S \approx 1.1 \times 10^4$ m/sec) and  BN ($v_S \approx 7 \times 10^3$ m/sec). Simulations of energy transport in Si$_3$N$_4$ and BN, using the same interaction potential, yield velocities $v_S \approx 7.5 \times 10^3$ m/sec and $v_S \approx 8 \times 10^3$ m/sec, respectively\cite{Sethi02a}.

This transport time $\tau_S$ competes with the time needed to dissipate the energy throughout the system, mostly due to phonon-phonon-interactions. This dissipation time also can be estimated from the heat pulse experiment\cite{Sethi02a}, to yield $\tau_{diss} \approx 50$ fs. One should note that this value strongly increases with the degree of crystallinity in the system (similar observations were made for amorphous Se when compared with crystalline Se\cite{Oligschleger99a}).

One important application of this analysis is that we can simulate and measure the heat conductivity of a-Si$_3$B$_3$N$_7$ by generating a steady-state heat flow through the system. To achieve this, we perform the following simulation procedure\cite{Oligschleger99a}:
\begin{enumerate} 
\item Equilibrate the system at constant temperature for a sufficiently long time $\tau_{wait}$ (see below) to keep aging to a minimum for the remainder of the experiment. 
\item Increase the temperature along a layer of atoms located at $L/4$ in the simulation cell by $\Delta T = 0.1 \times T$, and decrease it be the same amount along a layer at $3L/4$. 
\item One now lets the system relax on the time scale of the dissipation time. By then a temperature gradient is established, and a constant flow of heat into and out of the system is present. 
\item We maintain this steady-state situation for an observation time $t_{obs} < \tau_{wait}$, and measure the average heat flow from the hot to the cold plate.
\item This information can be used as input into the standard phenomenological heat transport law, 
\begin{equation}
\kappa = \frac{|j|}{|\nabla T|} = \frac{\langle \Delta E \rangle}{A \Delta t} \cdot \frac{L/2}{2 \Delta T},
\end{equation}
 to deduce the heat conductivity $\kappa$, where $|j| = \langle \Delta E \rangle  / (A \Delta t)$ is the average heat flow through a cross section $A$, and $|\nabla T| = 4 \Delta T / L$ is the temperature gradient. 
\item In a final step, one repeats this procedure for larger simulation cells, in order to take the limit of $\kappa$ for infinite cell size ($L \rightarrow \infty$). 
\end{enumerate}
Figure \ref{kappa} shows the extrapolated heat conductivity as a function of temperature. We note that this value, $\kappa \approx 4$ W/mK, shows only a small variation with temperature\cite{Sethi02a}, similar to the results obtained for amorphous Se.\cite{Oligschleger99a} Since no experimental value has been measured up to now for a-Si$_3$B$_3$N$_7$, we estimate the reliability of the procedure by studying the corresponding crystalline binary systems. One finds that the experimental value exceeds the one from the simulations by about a factor $3 - 5$, similar to what we had found in the case of Selen\cite{Oligschleger99a}. Thus, one would conclude that the true value of $\kappa$ should be about $\kappa_{exp} \approx 15$ W/mK.
\begin{figure}
\centering
\includegraphics[width=0.5\textwidth,angle=-90]{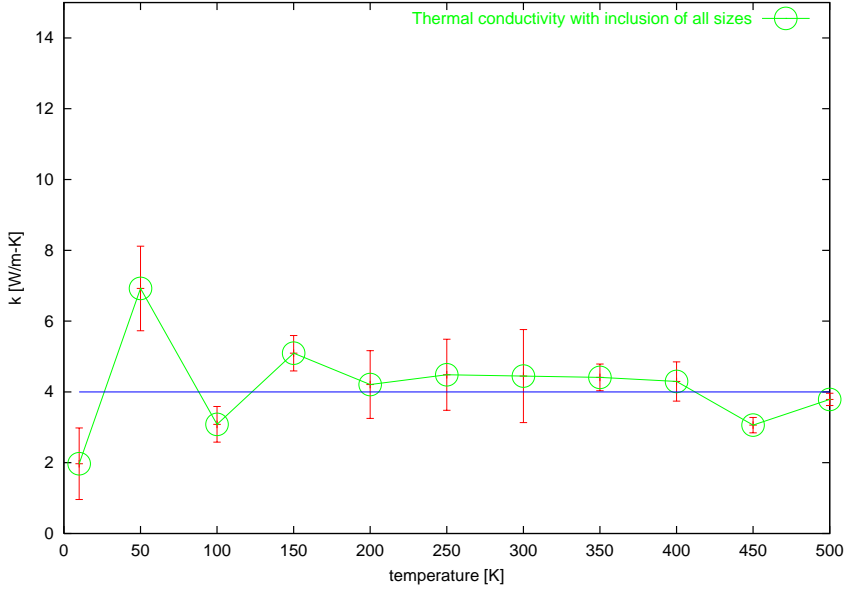}
\caption{Heat conductivity $\kappa$ as function of temperature $T$.
\label{kappa}} 
\end{figure}

\item Aging, Dissipation, and Response

Unless interested in relaxation processes as such, one usually waits until the system being simulated has relaxed into a state of partial or local equilibrium. Several time scales are of particular interest in this context: The time(s) $\tau_{init}$ necessary to "forget" various aspects of the initial conditions, and the time(s) $\tau_{resp}$ needed to react to the application of an external force. In the latter case, we might be dealing with either constant or time-varying, in particular sinusoidal, forces. We note that $\tau_{resp}$ usually implies a relatively weak external disturbance of a system that is already equilibrated. In contrast, $\tau_{init}$ refers to a relaxation from a possibly highly non-equilibrium initial condition of our simulations. 

Up to this point, these considerations are applicable to both crystalline and amorphous solids. But when simulating amorphous systems, one has to take into account that the region of the energy landscape we explore is not centered around a (global) minimum that is separated from similarly deep minima by insurmountable (on the time scale of the simulation) barriers. Instead, a permanent drift towards lower and lower energy minima takes place. Although the barriers needed to be crossed to reach the lower-lying minima constantly increase, this only leads to a logarithmic decrease in the speed with which deeper regions become accessible but does not stop this drift. 

As a consequence, we are dealing with a so-called aging process: For observation times short compared to the time $\tau_{wait}$ the system has been allowed to relax freely, the system behaves as if it were in thermal equilibrium, while for times exceeding $\tau_{wait}$ non-equilibrium behavior sets in. This can be seen by studying various two-time correlation functions, such as the two-time energy-energy average 
\begin{equation}
\phi(t_w, t_{obs};T) = \frac{\langle E(t_w) \cdot E(t_w+t_{obs})\rangle_{ens}}{\langle E(t_w)\rangle_{ens}^2}
\end{equation} 
or the closely related two-time autocorrelation function 
\begin{equation}
C_E(t_w,t_{obs};T)\equiv \langle E(t_w) \cdot E(t_w+t_{obs})\rangle_{ens}- \langle E(t_w)\rangle_{ens}\cdot\langle E(t_w+t_{obs})\rangle_{ens},
\end{equation} 
which in a (quasi)-equilibrium regime equals a generalized standard equilibrium specific heat $k_BT^2C_V(t_w,t_{obs};T)$.

In figure \ref{epot}, we show the potential energy $E_{pot}$ of local minima along MC-trajectories as function of time $\ln t_{obs}$ for several temperatures averaged over many trajectories\cite{Hannemann02b}. 
\begin{figure}
\centering
\includegraphics[width=0.6\textwidth,angle=0]{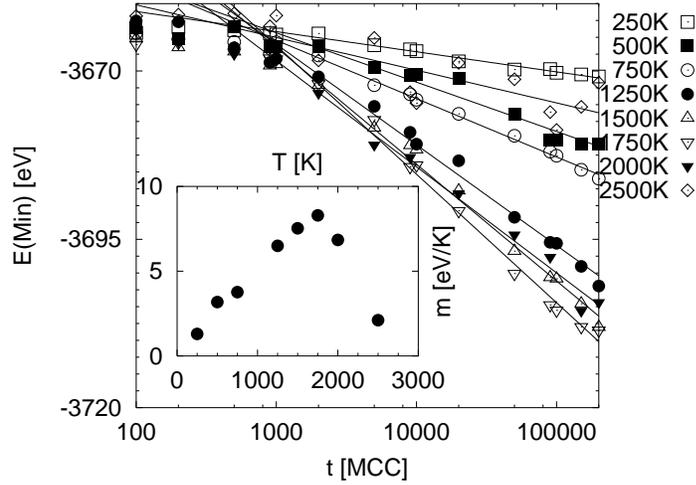}
\caption{Average potential energy $E_{pot}$ of local minima along trajectories as function of time $\ln t_{obs}$ for several temperatures. The inset shows the negative slopes $m$ as function of temperature. 
\label{epot}} 
\end{figure}

One clearly notices that after $\tau_{init} \approx 10^3$ MCC (Monte-Carlo cycles) the behavior changes and a clear downward drift of the energy occurs which is proportional to $\ln t$. This value of $\tau_{init}$ is also observed in other quantities such as the diffusion of atoms. For $t > \tau_{init}$, the 'ballistic' phase appears to be over, and vibrations about local minima together with slow hopping from minimum to minimum on the landscape best describe the dynamics in configuration space.

This drift on the landscape leads to aging as can be seen in figure \ref{aging1}, where we show the development of the two-time energy-energy average as a function of waiting time\cite{Hannemann02a}. The departure from the (quasi-)equilibrium value ($= 1$) for $t_{obs} > \tau_{wait}$ is clearly observable.
\begin{figure}
\centering
\includegraphics[width=0.6\textwidth,angle=-90]{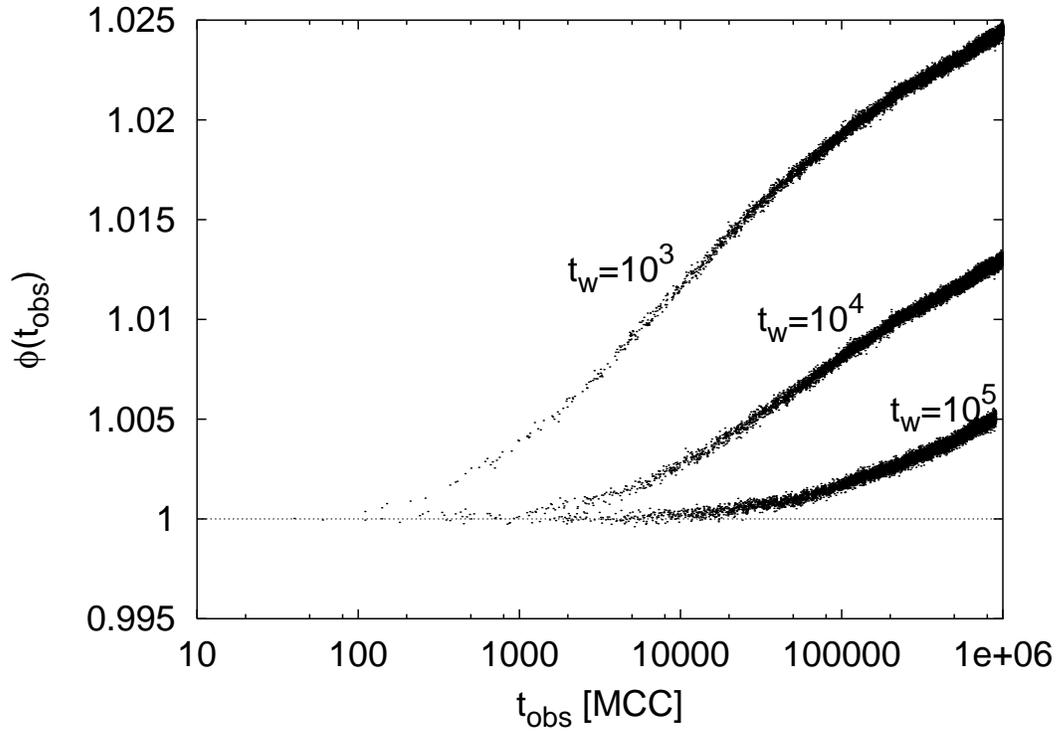}
\caption{Two-time energy-energy average $\phi(t_w, t_{obs};T)$ as function of time for several waiting times at $T=1250$ K for an ensemble of 100 runs.
\label{aging1}} 
\end{figure}

\item Change of the Network

The major events that take place during the aging process are associated with changes in the topology of the network. One quantity that reflects this behavior is the bond survival probability $BSP$, which is defined as the probability as a function of time that a bond that was present at time $t_1$ is still present at time $t_2$ (without having been broken in-between). Figure \ref{BSP} for the 
Si-N bonds shows that for high temperatures, where the system is essentially fluid, $BSP(\tau_{init},t)$ follows a stretched exponential curve, $ BSP(t) \propto \exp(-(\frac{t}{\tau_{BSP}})^{\beta})$; the curves for B-N bonds exhibit the same functional behavior. A similar behavior is found for low temperatures, although here one cannot observe the decrease to zero - on the time scale of the simulations, most of the bonds remained unbroken.
\begin{figure}
\centering
\includegraphics[width=0.53\textwidth,angle=-90]{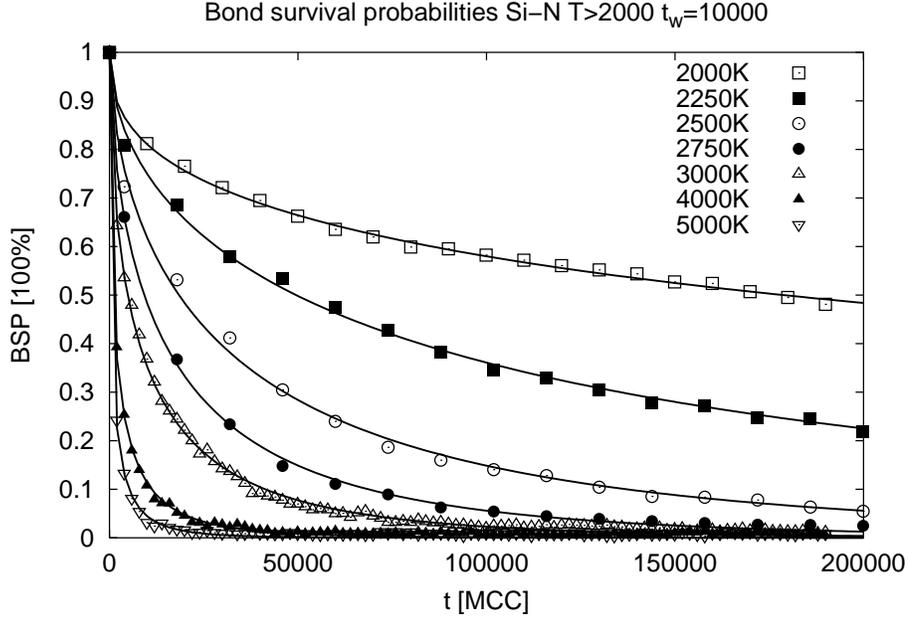}
\caption{Bond survival probability BSP as function of time for several temperatures.
\label{BSP}} 
\end{figure}

The characteristic time scale is $\tau_{BSP}$, which increases exponentially once we go below a critical temperature $T_c \approx 2000 - 2500$ K. Quite generally, the behavior of a-Si$_3$B$_3$N$_7$ is comparable to that of a glass when crossing this temperature region: A freezing-in of the structure, associated with a peak in the specific heat and an exponential increase of the viscosity\cite{Hannemann02a}. For temperatures below $T_c$, we observe some weak aging in the bond survival probability, i.e.\ the network becomes more resistant to change as time passes. This is not surprising considering the downward drift in the potential energy.

Nevertheless, the existence of a typical time scale for bond-breaking allows us to apply separation of time scales arguments to the modeling of the system. For $t_{obs} \ge \tau_{BSP}$ ($\tau_{BSP} \ge 10^{-7}$ sec for $T \le 500$ K), we can dispense of the vibrational degrees of freedom, and choose the topology of the system as the relevant quantity. It is then possible, to replace the continuous configuration space by a metagraph, where each node corresponds to a (possibly stretched) network topology, and the edges of this graph correspond to the possible bond-breaking and bond-stretching moves of the atoms. In order to assign realistic energies to these configurations, one would place the network on a lattice with an appropriately chosen lattice-spacing, and study the resulting energy landscape, and model the dynamics as a stochastic process on the metagraph.

Such work has been performed in the past for a number of discrete complex systems, such as the TSP-landscape\cite{Sibani93a}, chain-polymers\cite{Schoen02c} or one-component 2d- and 3d-lattice networks\cite{Schoen98a,Schoen00a}. While such calculations have not yet been attempted for a-Si$_3$B$_3$N$_7$, it would be a natural further step to take, in order to model dynamical properties of this system for very long times.
\end{itemize}

\section{Homogeneous Sintering}

Once we leave the realm of physical phenomena that can be described on picosecond time scales, we are even more urgently in need of physically reasonable simplifications, if we want to model and understand the various processes taking place during the generation of a macroscopic material, and its properties on macroscopic time scales. As an example we are going to analyze homogeneous sintering, and show how one can use a stepping stone approach to model the complete sintering process starting with clusters and ending with a dense macroscopic sinter.\cite{Salamon93a,Salamon93b}

We begin with a short summary of the sintering process, where for simplicity it is assumed that only one type of (metal) atoms is involved. Typically, at the outset a dense packing of small particles or atom clusters is prepared, which in many cases has already been arranged in a particular macroscopic shape needed as a part in a work-piece. This conglomerate is now heated to about $2/3$ of the melting temperature of the bulk material, leading to an enhanced diffusional motion of the atoms. The total amount of 'liquid' (= mobile atoms) present is relatively small, and the necks between the clusters are wetted. During this time, new 'liquid' will be produced throughout the sinter at places favoring high atomic mobility (e.g.\ naked surface of small spheres), while some of the 'liquid' will solidify in thermodynamic sinks of the material (e.g.\ in the interior of the necks connecting the spheres). At the beginning, the communication among the various 'liquid-like' wetted regions is rather slow, and the shapes of the neck-configurations correspond to locally equilibrated arrangements of the mobile atoms.

The surface tension generated by the wetting process leads to a net-movement of the clusters towards each other, in order to minimize the total surface energy, and a redistribution of the atoms at the necks and on the surface of the clusters. This process competes with the exchange of 'liquid' material among the various pores and necks, which would lead to a global equilibration among the wetted regions. At this stage, an important role can be played by the heat-flow through the system. This flow is rather complicated due to the many melting and re-freezing processes that can, in principle, take place within the sinter.

Once a certain densification of the solid-like skeleton has been achieved, the open channels providing for easy surface diffusion-driven communication among the 'liquid-like' regions are closed off, and atom transport is now only possible via grain boundary diffusion or bulk diffusion. At this stage of the process, it is energetically favorable for the isolated pores to become more spherical. Furthermore the annealing of the grain boundaries that had been generated at the necks between the spheres and along the closed off channels begins. In addition, the smaller pores start drifting, and mergers among them can occur at this stage of the process.

These processes continue into the final stage of sintering, where a coarsening process among the surviving (spherical) pores takes place. Here, atoms are exchanged via bulk and grain boundary diffusion, leading to a net growth of larger pores at the expense of smaller ones. In the ideal case, no pores will be left after a very long time, all "vacancies" having diffused to the surface. During the whole process outlined, the sinter is likely to have shrunk in volume by 20 - 40 \%. Controlling the local macroscopic shrinkage rates is therefore an important issue in the efficient production of pre-shaped work-pieces using the sintering process.

When trying to model this complicated process, preferably starting at the atomic scale, separation of time scales arguments turn out to be very helpful. Some of the relevant time scales for sintering are:

\begin{itemize}
\item Relaxation of electronic degrees of freedom ($\tau^{el} < 10^{-15}$ sec)
\item Lattice vibrations ($10^{-9} > \tau^{vib} > 10^{-14}$ sec)
\item Achieving local (spatial) equilibrium during wetting
($10^{-4} > \tau^{eq} > 10^{-8}$ sec)
\item Production of the mobile pool ($10^{-2} > \tau^{pr} > 10^{-6}$ sec)
\item Atomic transport (communication time) among locally equilibrated wetted regions ($10^{-2} > \tau^{com} > 10^{-6}$ sec)
\item Equilibration among wetted regions  ($10^{-2} > \tau^{eqens} > 10^{-6}$ sec)
\item Movement of solid particles across several atom-atom distances
($10^{-2} > \tau^{core} > 10^{-6}$ sec)
\item Annealing (movement) of grain boundaries
($10^{5} > \tau^{gb} > 10^{-1}$ sec)
\item Rounding of pores ($10^{3} > \tau^{ro} > 10^{-2}$ sec)
\item Equilibration of the atoms within grain boundaries ($10^{3} > \tau^{eqgb} > 10^{-4}$ sec)
\item Time needed for atoms to diffuse between isolated pores ($10^{5} > \tau^{diff} > 10^{-3}$ sec)
\end{itemize}
Of course, all these times are strongly dependent on pressure, temperature, connectivity of the sinter, type of atoms, etc.

These time scales suggest that modeling of sintering involves two different classes of models, the analysis of the atomic movement in an effective potential, and a continuum thermodynamic model that employs the parameters deduced from the atomic simulations as input. The first class includes the computation of parameters for an effective potential of atom movements, followed by molecular dynamics simulations of the behavior of clusters and their fusion. Finally it involves the calculation or estimation of various parameters and time scales needed in the thermodynamic model, such as diffusion constants, the time scales for movements of grain boundaries, the speed with which pores are closed off and become spherical, and the time scale for isolated pores to diffuse through the sinter and merge with other pores.

Here, we are going to focus on the thermodynamic continuum model, where we use the separation of time scales to identify the following processes:
\begin{itemize}
\item Fastest process: local wetting of pores and necks.
\item Slower competing processes: melting / re-freezing / transport of mobile material, equilibration among the wetted regions, movement of solid nano-particles constituting the solid skeleton.
\item Slowest processes: annealing of grain boundaries, drift of isolated pores, and coarsening of isolated pores via emission and absorption of atoms / vacancies.
\end{itemize}

This general concept was implemented with several groups participating. J. Bernholc and co-workers at NCSU derived EAM-potentials\cite{Bernholc93a}, and H.P. Cheng and R.S. Berry in Chicago studied the melting and fusion of clusters\cite{Cheng91a,Cheng92a}. Summarizing their most important results, they observed surface melting for clusters and found that a neck grew between two clusters via wetting and that a grain boundary was created along the neck, which only slowly could be removed.

The process of sintering several clusters, through the closure of pores, their movement and merger, and coarsening via vacancy emission was studied using a very simplified lattice-gas type model, in order to be able to bridge the many time scales involved\cite{Schoen93a,Kjeldgaard94a}. It was found that the closure of pores did occur relatively fast\cite{Schoen93a}, and that the shape of the clusters and pores exhibited strong fluctuations about the spherical "ideal" minimum energy shape. For small pores, merging was the dominant coarsening process for high pore density (the density of the pores was essentially given by the overall density of the sinter). In contrast, for larger pores, standard Ostwald-ripening occurred except for very closely neighboring pores, where large shape fluctations (e.g. a massive elongation of the pore) could lead to a direct contact among these pores resulting in a fast merger and rounding of the merged pore.\cite{Kjeldgaard94a}

Besides the separation of time scales, the guiding principle of the thermodynamic model is the minimization of the (free) energy. This energy minimization corresponds to the determination of the average values of the fast degrees of freedom in partial and/or local thermodynamic equilibrium. Regarding the initial configuration of our model of the sintering process, we note that the starting sinter is created by densely packing small particles, perhaps under application of pressure. Thus, a minimum energy configuration corresponds to an as dense as possible packing of these particles. Finding the densest packing for e.g.\ spheres of different sizes is a global optimization problem. But even determining local minima, i.e.\ arrangements that can no longer be compressed any further by infinitesimal shifts of the spheres and are stable against small to moderately large rearrangements of the spheres, is a non-trivial task. Such arrangements of spheres constitute random close packings of spheres, which we have generated by appropriate algorithms.\cite{Frost93a}

The next step in our approach requires the wetting of pores and necks. At the beginning, the optimal location for the mobile pool of atoms is within the necks between pairs of fixed unchangeable nearest-neighbor spheres. The minimum energy configuration is given by minimizing the surface energy of the two-sphere system for a given amount of liquid (Plateau's problem). While this task can be solved analytically in terms of elliptic functions, we have found it more useful to directly solve the corresponding Euler-Lagrange equation with the appropriate boundary conditions, and tabulate the shapes of the optimal solid+liquid configurations as a function of radii of the spheres, center-center distance of the spheres and amount of liquid available. For single necks, the minimum energy surface is axisymmetric, exhibits constant mean curvature, and the cross section is given by a Delaunay curve.\cite{Basa91a,Basa94a} 

The transport of liquid material between the wetted regions can be modelled via network thermodynamics.\cite{Schoen93b,Salamon93a} Here, one describes the net-diffusive transport $I_{ij}$ between many sources and sinks $i$ and $j$ containing $N_i$ and $N_j$ atoms, respectively, in analogy to an electric network consisting of resistors and capacitors. In the model, the conductivity and the voltage differences correspond to the effective diffusivity $\sigma_{ij}$ along the channel or grain boundary and the difference in chemical potentials $\mu_j(N_j) - \mu_i(N_i)$ (surface energies of the wetted regions), respectively:
\begin{equation}
\frac{dN_i}{dt} = \sum_{j \ne i}I_{ij} = \sum_{j \ne i}\sigma_{ij}(\mu_j(N_j) - \mu_i(N_i))
\end{equation}
We note that this model also applies to the description of the interaction via grain boundary diffusion of a dense distribution of isolated pores at the beginning of the coarsening process.

The movement of the individual solid spheres follows simple Newton's equations, where three types of forces contribute:
\begin{itemize}
\item Force, due to minimization of local surface tension: 
\begin{equation}
\vec{F}_{s} = -\sigma \sum_{i=1}^{n}\frac{\partial A_i(|\vec{h}_i|)}{\partial h_i}\hat{\vec{h}}_i,
\end{equation}
where $\hat{\vec{h}}_i$ is the unit vector along the center-center axis between neighbor spheres along neck $i$, $\sigma$ the surface tension, and $A_i(|\vec{h}_i|)$ the surface area of the neck, respectively.
\item Damping force due to the viscosity caused by refreezing of the necks: 
\begin{equation}
\vec{F}_{vis} = -\eta \vec{v}_i,
\end{equation}
where $\eta$ is the viscosity.
\item Forces due to steric constraints.
\end{itemize}
Depending on temperature, viscous forces or steric constraints tend to dominate, in the sense that we can assume that the particles either move with a nearly constant ('final') velocity (dominated by viscous forces) or move at each step the maximum distance allowed until solid-solid contact is established.\cite{Schoen91a,Salamon93a,Salamon93b}

If the changes in the skeleton are slow compared to the transport of material among the liquid pools and the production of new liquid material via surface melting of the spheres, the network thermodynamics model applies. In the extreme case of very slow addition of new material to the liquid pool (low rate of surface melting) and nearly unchanged solid skeleton, then a global equilibration among the wetted regions can take place. This equilibration constitutes again a global optimization problem\cite{Schoen92a}, and can be solved e.g.\ using dynamical programming\cite{Schoen91a}.

On the other hand, if the liquid production is very slow, but liquid transport is fast and the solid particles easily movable, we are dealing with a generalization of Plateau's problem, i.e., we search for the minimum surface of a combination of many solid movable particles plus a freely deformable liquid pool\cite{Schoen92b}.

The most cumbersome task we are faced with is the determination of the detailed shape of the wetted regions once several necks coalesce, but the channels between the pores have not yet closed. No analytical solutions are available (the solutions are no longer axisymmetric surfaces), and the number of possible multi-sphere+liquid configurations that need to be considered becomes very large. Therefore, tabulating the numerical solutions of the Euler-Lagrange equations does no longer appear to be an efficient approach, and one has to use e.g.\ a numerical minimum surface solver\cite{Brakke90a} at this stage after every liquid-redistribution step. However, our simple model simulations mentioned above indicate that this stage does not last very long, and that the pores rather quickly become isolated. Once this has happened, rounding of the pores begins to take place, together with a slow drift, both of which are amenable to analytical modeling. Depending on the size and the density of the pores, the interaction at this stage will be best modeled by network thermodynamics or by models of merging pores analogous to the von Smoluchowski-approach.\cite{Kjeldgaard94a}

Once the pores can be considered as essentially immovable, standard coarsening processes will take over. Here, one needs to address the question of bulk vs. grain boundary diffusion, and the issue of shielding of the pores by neighboring pores. Clearly, the models depend on whether the interaction among the pores proceeds via three- or two-dimensional diffusion. Bulk diffusion clearly goes via three dimensions, while the case of grain boundary diffusion is not as clear-cut. We might regain three-dimensional diffusion, if the grain boundaries intersect very frequently in 'zero volume' pores or lines. In such a case, the vacancies / atoms within the two-dimensional grain boundary network 'notice' the three-dimensionality of the embedding space, and the net-diffusion throughout the solid can be treated as if it were three-dimensional, although with a different effective diffusion coefficient.
 
If the grain boundary network is rather sparse with each grain boundary connecting many pores and most pores intersecting one grain boundary only, however, we can replace the three-dimensional sinter by a single two-dimensional plane that intersects all the three-dimensional pores.\cite{Salamon93a,Schoen92c} In this situation, we can solve the equations involved (2d-diffusion-, 3d-Fokker-Planck-equation, 3d-conservation of vacancies) employing some generalization of the standard procedures. 

Again, separation of time scales comes to our assistance by allowing us to separately analyze the process of rounding of the pores to spherical cavities, the establishment of a diffusional flow of one-atom-size vacancies between the individual pores and the environment at infinity, and also the net exchange of material among pores and the growth of the average size of the surviving pores. (Note that at the final stage the dynamics takes place on a metagraph, whose nodes correspond to the feasible distributions of clusters of various sizes, and whose edges correspond to allowed changes of the cluster distribution.)

The result is similar to the one derived for bulk-diffusion\cite{Lifshitz81a}, i.e.\ the scaled radius $x$ is related to the scaled time $t'$ via $x \propto (t')^{1/3}$, but with correction terms:
\begin{equation}
x = \left (\frac{4}{9}(t' - p_0)\right )^{1/3} \left [ 1- \frac{9}{8}(4/9)^{2/3}f_2 (t' - p_0)^(-1/3) + O(f_2^2(t'-p_0)^{-2/3})      \right ],
\end{equation}
where $p_0$ and $f_2$ are combinations of several constants characteristic of the sintering material and its state at the beginning of the coarsening process.

We note that in the standard approach, the pores only interact via a sea of vacancies (at infinity), either in the bulk (3d) or in the infinite grain boundary (2d). If the density of pores is still quite high, however, this can no longer be assumed to hold, and shielding of the pores by neighboring pores becomes important. Again, this problem can be treated analytically on various time scales\cite{Kjeldgaard94a}, and we find that e.g.\ the mean cluster size $\overline{M}$ in $d$ dimensions grows as 
\begin{equation}
\overline{M}(t) \propto t^{d/(4+d)}.
\end{equation}

At first sight, this example of a homogeneous sinter might appear to be far removed from the field of non-crystalline and partially crystalline materials, which has been the topic of this workshop. But from the point of view of the modeler both the general techniques employed and even many of the particular algorithms presented can easily be adapted to the analysis and modeling of macroscopic dynamical processes in amorphous systems, such as the growth of crystals from the amorphous phase or the generation of nanocrystalline material with a specific texture.

\newpage
\renewcommand{\baselinestretch}{1.0}
\normalsize
\bibliography{/c3serv1/schoen/Bibliography/notitle}
\renewcommand{\baselinestretch}{1.6}
\normalsize

\end{document}